\documentclass[lettersize,journal]{IEEEtran}
\usepackage{amsmath,amsfonts}
\usepackage{algorithmic}
\usepackage{algorithm}
\usepackage{array}
\usepackage[caption=false,font=normalsize,labelfont=sf,textfont=sf]{subfig}
\usepackage{textcomp}
\usepackage{stfloats}
\usepackage{url}
\usepackage{verbatim}
\usepackage{graphicx}
\usepackage{cite}
\usepackage{footmisc}

\usepackage{xcolor}
\usepackage{color,colortbl}
\usepackage{verbatim}
\DeclareMathOperator*{\sign}{\mathrm{sign}}

\newcommand\smallbullet{\scalebox{0.65}{$\bullet$}}%

\usepackage{mathrsfs}

\newcommand\leth{\frac{\mathrm{d}t}{\mathrm{d}s}}

\usepackage{tikz}
\newcommand\copyrighttext{%
  \centering\footnotesize \textbf{\copyright 2024 The Authors}.  
  \textbf{Cite as:} G. Riva, S. Radrizzani and G. Panzani, "Battery model impact on time-optimal co-design for electric racing cars: review and application," in IEEE Transactions on Transportation Electrification, doi: 10.1109/TTE.2024.3403930}
\newcommand\copyrightnotice{%
\begin{tikzpicture}[remember picture,overlay]
\node[anchor=south,yshift=-2cm] at (current page.north) {\fbox{\parbox{\dimexpr0.9\textwidth-\fboxsep-\fboxrule\relax}{\copyrighttext}}};
\end{tikzpicture}%
}

\begin{document}
%%%%%%%%%%%%%%%%%%%%%%%%%%%%%%%%%%%%%%%%%%%%%%%%%%%%%%%%%%%%%%%%%%%%%%%%%%%%%%%%
% TITLE AND AUTHORS
%%%%%%%%%%%%%%%%%%%%%%%%%%%%%%%%%%%%%%%%%%%%%%%%%%%%%%%%%%%%%%%%%%%%%%%%%%%%%%%%
\title{
Battery model impact on time-optimal co-design for electric racing cars:
review and application
} 

\author{Giorgio Riva,~\IEEEmembership{Member,~IEEE,} Stefano Radrizzani,~\IEEEmembership{Member,~IEEE,} Giulio Panzani,~\IEEEmembership{Member,~IEEE}
\thanks{The authors are with Dipartimento di Elettronica, Informazione e Bioingegneria, Politecnico di Milano, Via G. Ponzio 34/5, 20133, Milan, Italy (e-mail: giorgio.riva@polimi.it, stefano.radrizzani@polimi.it, giulio.panzani@polimi.it). Corresponding author: Giorgio Riva.
 }}
 
\maketitle
\copyrightnotice
\begin{abstract}
The sustainable mobility trend touches the racing world as well, from the hybridization of Formula 1 (F1) and Le Mans Hypercars to the fully electric Formula E racing class. In this scenario, the research community is studying how to push electric racing vehicles to their limit, combining vehicle dynamics and energy management, to successfully solve the minimum lap time problem. 
Recently, this class of problems has been enlarged towards optimal sizing, with a particular interest in batteries, which represent the main bottleneck for electric vehicle performance. 
In this work, starting from a thorough review of literature approaches, we define a general optimization framework of minimum lap and race time problems for electric vehicles, suitable to figure out the impact of different modeling choices on both problem structure and optimal variables profiles.
Exploiting a case study on Generation 3 (Gen 3) of Formula E cars, we delve into the impact of battery models' complexity on both optimal sizing and optimal battery usage. We show how highly detailed models are necessary to study the evolution of both battery and vehicle control variables during the race, while, simple models are more than sufficient to address the battery sizing problem.
\end{abstract}

%%%%%%%%%%%%%%%%%%%%%%%%%%%%%%%%%%%%%%%%%%%%%%%%%%%%%%%%%%%%%%%%%%%%%%%%%%%%%%%%
% INTRODUCTION
%%%%%%%%%%%%%%%%%%%%%%%%%%%%%%%%%%%%%%%%%%%%%%%%%%%%%%%%%%%%%%%%%%%%%%%%%%%%%%%%
\section{Introduction}
In the last decade, the interest in vehicles equipped with electric powertrains has considerably increased in light of the transition towards sustainable mobility. In this context, many technological challenges have emerged \cite{sanguesa2021review}, and batteries represent one of the most impelling obstacles towards the full-scale adoption of Electric Vehicles (EV). Indeed, the current technologies available on the market suffer from multiple weaknesses, such as high costs, lower energy and power densities than fuel, and safety concerns \cite{zeng2019commercialization}.

As often happens in the automotive field, the most challenging technological progresses are driven by racing competitions, and vehicle electrification does not represent an exception. This fact is demonstrated by the hybridization pursued by Formula 1 (F1) and Le Mans Hypercars, and, most importantly, by the fully electric Formula E racing class, which has just revealed the plan for the new Generation 4, which will start racing in 2026.
In racing competitions, performance is merely evaluated by the lap time, 
and in the specific case of electric vehicles, batteries represent the main bottleneck for performance. Indeed, manufacturers have to cope with the trade-off between the high energy and high power density of current state-of-the-art technologies, as a fragile balance among disposable energy, power, and weight. The pivotal role of batteries is also motivated by the continuous development of new technologies for automotive (see \cite{radrizzani2023optimal} for a complete overview), which represents a significant opportunity in the direction of performance enhancement in the future.
Compared to urban vehicles, where the sizing of batteries and other powertrain components is chosen to maximize mileage and durability while satisfying economical constraints, in the racing context the focus is shifted towards the maximization of the vehicle performance given the requirements of the specific competition, e.g., mileage, weight, and power limitations.
Thus, in the described scenario, the ability to evaluate the performance and to select the size of powertrain and battery pack components, for a given application, represents a key tool for racing vehicle manufacturers.

\textbf{\textit{Literature Review.}}
As a consequence, in the scientific literature many works have emerged tackling the problem of performance optimization for electric racing vehicles, where the role of the different components is taken into account, as in \cite{borsboom2021convex,yu2018codesign}, with an eye on batteries in \cite{radrizzani2023optimal,riva2022sizing}. These works rely on the definition of an Optimal Control Problem (OCP), where each vehicle component is suitably modeled, and the time to cover a single lap or a full race is minimized, as a proxy for the vehicle performance.
Minimum Lap Time (MLT) problems are widely studied in the scientific community of vehicle dynamics, as a tool to analyze and understand how to improve vehicle performance \cite{perantoni2014optimal,massaro2021minimum}, from the coupling of Quasi Steady State (QSS) methods and g-g plots \cite{rice1973measuring}, up to modern OCP solutions \cite{de2022minimum}, starting from the 1990s.
The family of OCP consists of a nonlinear optimization problem, setup to minimize the lap time and constrained by the vehicle dynamics. In this context, employed models span a variety of approaches: the vehicle can be modeled through either a simple point mass or a more complex multi-degrees-of-freedom (DOF) model, e.g., in \cite{limebeer2015optimal,dal2018minimum}, while the trajectory can either be dictated or left free as an additional design variable. 
In the world of hybrid and electric vehicles, the standard vehicle model
is the single DOF point mass driving on a fixed trajectory (see \cite{borsboom2021convex,ebbesen2018minlap, salazar2018equivalent, salazar2019minlap, duhr2021time,locatello2021time, duhr2022convex, duhr2023minimum}), especially when the focus is on the design of either the energy storage system or the powertrain, since it is able to grab the most relevant energy-related phenomena. As exceptions, \cite{broere20224wheel} and \cite{yu2018codesign} propose MLT solutions employing respectively a double track model with load transfer, and a fourteen DOF model including suspensions.
The literature of OCP for hybrid electric vehicles is polarized towards the convex formulation based on \cite{murgovski2015convex}, from the first work in \cite{ebbesen2018minlap} on a hybrid F1 car, followed by \cite{salazar2018equivalent} and \cite{salazar2019minlap}.
Inside this convex framework, \cite{borsboom2021convex} represents the first example of MLT for fully electric racing vehicles, where special effort is given to model and analyze the impact of both electric motor and transmission layouts. 
In all mentioned works, given the lap time objective, the battery is always modeled through a highly simplified equivalent circuit with constant voltage.
In the light of sizing and management of the energy storage system, together with the temperature management of powertrain elements, a natural extension of MLT problems is represented by Minimum Race Time (MRT) ones, where mileage represents a key aspect in the optimization.
In this scenario, \cite{herrmann2020minimum} proposes a trajectory optimization OCP to devise a two laps race strategy for an autonomous electric vehicle, accounting for the temperature of the electric motor; while, in the flavor of convex formulations, \cite{vankampen2022maximumdistance} enlarges the MLT problem to a race as a sequence of stints for an endurance car, to outcome the optimal race strategy. 
In this context, the optimal sizing of batteries for electric racing vehicles has been addressed only by \cite{riva2022sizing} and by its hybrid architecture extension in \cite{radrizzani2023optimal}, which have first introduced a more complex battery model, where the voltage depends on the state-of-charge of the battery, to retrieve more reliable battery outcomes. These two works tackle the race problem in the same way as MLT ones, enlarging the optimization to the whole race length. 
It is worth mentioning that, different approaches to the race problem are proposed in \cite{vankampen2022maximumdistance} and \cite{duhr2023minimum}, where a model of the lap time is learned from a large multitude of MLT-OCP solutions (e.g., through Artificial Neural Networks as in \cite{duhr2023minimum}), and the whole race is optimized through a simpler higher level problem, merging multiple laps. 

\textbf{\textit{Aim and Scope.}} Considering the described state-of-the-art, the objective of our research is the study of the impact that different models can have on the structure of the resulting OCP, and on the obtained optimal size of a battery pack. In particular, we aim at highlighting the role of battery models, whose variety of choices entails the need for a precise guideline to catch different design objectives. In this way, we can address the problem of understating the amount of battery information that is needed a-priori when designing the battery pack for a racing application. Moreover, as a byproduct, we aim to tidy up the multiplicity of literature approaches for each component of a racing EV, inside a unified mathematical framework.

\textbf{\textit{Main Contributions.}} The novel contributions of our work can be decomposed into two main branches.\\
(i)
We outcome a comprehensive literature review of minimum lap-race time problems for electric racing vehicles, with a focus on the different models employed for each subsystem: vehicle dynamics, powertrain and brakes, and batteries.
Moreover, we show that the manifold of approaches available in the literature can be framed in a unified mathematical description of the problem, whose structural breakdown allows a straightforward analysis of its properties as a function of the modeling choices.
Finally, to enhance the overall readability, all the derived model formulations are summarized in Tab. \ref{tab_formulation_recap}, where the entire OCP is reconstructed, both in convex and non-convex forms.\\
(ii) Regarding the impact of battery models, we leverage the previous works in \cite{riva2022sizing} and \cite{radrizzani2023optimal} as a starting point, moving the focus from the selection of the best chemistry towards the choice of the most suitable battery model as a function of the designer objective. In this respect, we derive a practical guideline to select the complexity of the battery model, tailored to the desired objective, via a meaningful case study: the Formula E Generation 3 (Gen 3) in the 2021 Rome ePrix circuit, i.e, the current most challenging racing scenario for fully electric vehicles. The results suggest that the simplest possible battery model, whose parameters can be retrieved from easy-access datasheets, is sufficient to attain consistent sizing results, while the need for realistic behaviors of the optimized variables, for control purposes, can be obtained only via the more complex models, whose characterization might ask for further identification processes.
In addition, we provide a detailed comparison between convex and non-convex formulations, showing their consistency in terms of race time and sizing results, but also their differences in the optimal profiles introduced by the needed relaxations in the convex solutions. 

\textbf{\textit{Outline.}} In Section \ref{sec:prb_formulation}, the joint minimum race time and battery sizing problem is formally defined, highlighting the structural breakdown of the constraints and deriving its bi-level formulation. In Section \ref{sec:modelling}, the different modeling approaches of each vehicle subsystem are analyzed in detail, discussing the impact of the different choices on the convexity property of the optimization problem. Finally, the case study is tackled in Section \ref{sec:comparison}, where the impact of the derived battery models is discussed both in terms of sizing results and of the obtained profiles reliability, providing a useful guideline for the problem.
The paper is ended in Section \ref{sec:conclusions} with concluding remarks.

%% SECTION 2
\section{The Joint Minimum Race Time and Battery Sizing Problem}
\label{sec:prb_formulation}
As already discussed in \cite{riva2022sizing}, the joint optimization of race time and battery size can be cast as a single Co-design Optimization Problem (Co-OP), relating the sizing parameters of the battery pack to their impact on the achievable race performance on a given circuit.
In the literature, this co-design framework represents a common approach to tackle sizing problems of hybrid electric vehicles, both in racing, as in this work, and in non-racing applications. In the latter scenario, the final objective is typically different from time minimization (see \cite{silvas2017review, hu2016integrated, sundstrom2010torque}), and mainly focused on standard hybrid powertrains. In this work, we build up the formulation provided in \cite{riva2022sizing} for the racing context, highlighting how different modeling choices, with special attention to batteries, impact the properties and the outcomes of the resulting Co-OP.
Towards this aim, we structurally divide the set of optimization variables $\mathcal{U}$ of the Co-OP as $\mathcal{U}=[u_s,u_v,u_p,u_b]$, where $u_s$ represents the set of sizing battery parameters for a fixed cell technology, and
$u_v$, $u_p$, $u_b$ collect, respectively, all state and control variables related to the vehicle dynamics, the powertrain and braking system, and the battery.
From this description, the Co-OP can be mathematically formulated as:
\begin{subequations}
\label{eq_min_general}
\begin{align}
\displaystyle \min_{\mathcal{U}= [u_s,u_v,u_p,u_b]}t_{\mathrm{race}}(\mathcal{U}) \label{eq:prb1_cf} \\ 
u_s \in \mathscr{S} \label{eq:prb1_con_size}\\
\left[u_v,u_s\right] \in \mathscr{V}\label{eq:prb1_con_veh}\\
\left[u_p,u_s\right] \in \mathscr{P} \label{eq:prb1_con_pt}\\
\left[u_b,u_s\right] \in \mathscr{B} \label{eq:prb1_con_batt}\\ 
\left[u_v,u_p,u_b,u_s\right]\in \mathscr{C}, \label{eq:prb1_con_coupling}
\end{align}
\end{subequations}
where, the objective $t_\mathrm{race}$ defined in \eqref{eq:prb1_cf} represents the overall race time to be minimized, while the feasible set, defined by the intersection of constraints \eqref{eq:prb1_con_size}-\eqref{eq:prb1_con_coupling}, can be easily interpreted, thanks to the partition of $\mathcal{U}$ as follows.
\begin{enumerate}
    \item[(i)] Set $\mathscr{S}$ defines the admissible values for the tunable battery size parameters, which are typically expressed through box constraints. 
    \item [(ii)]
    Sets $\mathscr{V},\mathscr{P},\mathscr{B}$ model the evolution of the state variables and the possible physical limitations respectively of vehicle dynamics, powertrain and battery subsystems. Such three constraint sets are independent of the other subsystems' variables, while they can be affected by the sizing ones $u_s$, as shown in Section \ref{sec:modelling}. As an example, $\mathscr{B}$ constrains the evolution of the state-of-charge of the battery based on both the battery current and the size of the battery itself.
    \item[(iii)] 
    Set $\mathscr{C}$ models the coupling among the subsystems, and possibly the sizing variables.
\end{enumerate}

As already discussed in \cite{riva2022sizing} and \cite{fathy2001coupling}, Problem \eqref{eq_min_general} can be reformulated, safeguarding the global optimum, as a bi-level optimization problem composed of two nested layers: the sizing variables $u_s$  are optimized in the outer layer, where the cost function is evaluated through the inner one, which solves the minimum race time problem, given a fixed value for the sizing variables $\bar{u}_s\in \mathscr{S}$. The described bi-level optimization problem can be formulated as:
\begin{equation}
\displaystyle \min_{u_s \in \mathscr{S}}
\left\{
\begin{array}{c}
\begin{aligned}
     \displaystyle \min_{u_v,u_p,u_b} t_{\mathrm{race}}(u_v,u_p,u_b,u_s) \\
    u_v \in \mathscr{V}(u_s) \\
    u_p \in \mathscr{P}(u_s) \\
    u_b \in \mathscr{B}(u_s) \\ 
    \left[ u_v,u_p,u_b \right]\in \mathscr{C}(u_s)
\end{aligned}
\end{array}
\right\}.
\label{eq_two_stage}
\end{equation}

Before moving on, we introduce here a fundamental modeling tool, common to all subsystems of the problem. As explained before, the main objective of the sizing problem in \eqref{eq_min_general} for a racing vehicle is the minimization of the race time, which can be expressed as:
\begin{equation}
    \displaystyle
    t^*_{race}
    = \arg \min \,\,t(S),
    \label{eq_cf}
\end{equation}
where $t$ is the time, $S$ is the race length. 
To formulate an Optimal Control Problem (OCP), we exploit a state-space reformulation where the covered space along the race-line $s$ becomes the independent variable \cite{ebbesen2018minlap},\cite{versch2016} and the link between space and time is given by the vehicle speed $v$:
\begin{equation}
   v = \frac{\mathrm{d}s}{\mathrm{d}t} \rightarrow  \frac{\mathrm{d}\smallbullet}{\mathrm{d}s} = \frac{1}{v}\cdot\frac{\mathrm{d}\smallbullet}{\mathrm{d}t}.
   \label{eq_spt_ref}
\end{equation}
As extensively used in the following section, \eqref{eq_spt_ref} represents the building block to transform all dynamic equations in each subsystem of Problem \eqref{eq_two_stage} from time to space domain.
To highlight the strong dependence of all state equations from the vehicle speed $v$, Problem \eqref{eq_two_stage} can be  rewritten as:

\begin{equation}
\displaystyle \min_{u_s \in \mathscr{S}}
\left\{
\begin{array}{c}
\begin{aligned}
    \displaystyle \displaystyle \min_{u_v,u_p,u_b} \int^{S}_{0} \frac{1}{v(s)} \mathrm{d}s  \\
    %s.t.:\\
    u_v \in \mathscr{V}(u_s) \\
    \left[u_p,v\right] \in \mathscr{P}(u_s) \\
    \left[u_b,v\right] \in \mathscr{B}(u_s) \\ 
     \left[u_v,u_p,u_b\right]\in \mathscr{C}(u_s)
     \end{aligned}
\end{array}
\right\},
\label{eq_two_stage_speed}
\end{equation}
where $v$ is omitted in the first constraint because, as it will be explained in the next section, it always belongs to vector $u_v$. Thus, we remark that the vehicle speed profile is an optimization variable, and is not imposed a-priori.
The formulation of Problem \eqref{eq_two_stage_speed} provides a plain description of the optimization problem, where each subsystem -- vehicle, powertrain and battery -- is modeled through an independent set of constraints plus the coupling terms, possibly shaped by the fixed sizing variables. 

In the following sections, a comprehensive overview of the modeling approaches employed in the literature for each subsystem is presented, highlighting their impact on the structure of the optimization problem, and on the significance of both sizing results and optimal profiles via a suitable case study. 

\section{Minimum Race Time Modeling}
\label{sec:modelling}
As anticipated, the focus of this work is the analysis of the properties of the inner layer problem -- i.e., the minimum race time one -- looking at the impact on the sizing results of the outer one. So, we remark that we do not deepen the properties of the outer layer, mainly for two reasons: (i) it represents a different topic that requires a dedicated analysis, and (ii) in most cases, like in the example of Section \ref{sec:comparison}, it can be easily solved either through greedy brute-force approaches, like gridding, or via more efficient gradient-free solutions, as in \cite{radrizzani2023optimal,riva2022sizing}, with Bayesian Optimization, due the limited size and discrete nature of the problem. 

Given this premise, in this section, the different modeling approaches employed in the literature for each subsystem of the inner layer of Problem \eqref{eq_two_stage_speed} are considered, analyzing their impact on the problem structure. A specific focus is given to the \textit{convexity} property, which plays a fundamental role in optimization \cite{bertsekas2015convex}. Indeed, convex formulations of optimization problems can be very convenient for two main reasons \cite{bertsekas2015convex}: (i) any local minimum is also a global one, and (ii) the availability of reliable and efficient off-the-shelf solvers. Thus, convexity represents a nice to have property also for the class of problems tackled in this work, in order to guarantee global optimality and ease of solution despite the possible huge dimension of the problem itself, as discussed by many works, like \cite{ebbesen2018minlap,borsboom2021convex}. On the other hand, many physical laws employed in these problems are modeled by non-convex relationships, requiring possible significant approximations to keep them convex. This clean trade-off between convex and non-convex approaches is investigated in detail in this section from a modeling prospective, while the effect on results is addressed in Section \ref{sec:comparison}. 
We refer to Tab. \ref{tab_formulation_recap} for a summary of all the possible models described in the following of the section.\\

% Vehicle Dynamics
\subsection{Vehicle Dynamics Model}
\label{sub_sec_veh}
In the minimum lap/race time framework, especially when the main goal is the sizing of the energy storage system, the point-mass model is typically employed \cite{borsboom2021convex,ebbesen2018minlap, salazar2018equivalent, salazar2019minlap,duhr2021time, locatello2021time, duhr2022convex, duhr2023minimum}, to face the trade-off between model accuracy and complexity. The reason is two-fold: (i) the race-line is considered fixed and lap invariant, and (ii) the point-mass model is sufficient to capture all the energy related phenomena, which are the key elements for the battery sizing problem, as validated from an energy-related perspective in \cite{riva2022sizing} or \cite{duhr2022convex} against a commercial simulator.
The point mass model describes the dynamics of a vehicle only by means of the evolution of the longitudinal speed, which is computed by the following force balance:
\begin{equation}
    \frac{\mathrm{d}v}{\mathrm{d}t} = \frac{1}{M}\left( \frac{T_w}{R_w} - F_\mathrm{drag}(v)- F_\mathrm{roll}(v,\theta)  - F_g(\theta) \right),
    \label{eq_long}
\end{equation}
initialized as:
\begin{equation}
    v(0)=v_0,
    \label{eq_initial_condition}
\end{equation}
where $v$ is the vehicle speed
tangential to the fixed race-line, which is described by the curvature $\rho_R(s)$ and slope $\theta(s)$.  The force $F_\mathrm{drag} = C_\mathrm{drag}v^2$ represents the drag resistance, defined by 
the aerodynamic drag coefficient $C_\mathrm{drag}$, which is the predominant friction contribution in racing cars due to the high speed. The force $F_\mathrm{roll}=C_\mathrm{roll}F_z$ is the rolling resistance, which is proportional, via the rolling coefficient $C_\mathrm{roll}$, to the vertical load $F_z$, which is defined later in \eqref{eq_forces}. Force $F_g = Mg\sin(\theta)$, instead, is the gravitational resistance, where $g = 9.81$ m/s$^{2}$ is the gravitational acceleration. Parameter $M$ is the total mass of the vehicle, comprising the battery $M_b$ and the chassis $M_v$ one, and $R_w$ is the radius of a single equivalent wheel. Finally, $T_w$, i.e., the total torque applied to the wheel, is the control variable that affects the longitudinal speed, while $v_0$ represents the initial speed, which can model either a standstill or a rolling start.
Following the state-space reformulation introduced in \eqref{eq_spt_ref}, the vehicle state equation becomes:
\begin{equation}
\label{eq_veh_model_space_ref}
\resizebox{1.0\hsize}{!}{$
    \frac{\mathrm{d}v(s)}{\mathrm{d}s} = \frac{1}{M}\left( \frac{T_w(s)}{R_w} - F_\mathrm{drag}(v(s))-F_\mathrm{roll}(v(s),\theta(s))  - F_g(\theta(s)) \right)\cdot \frac{1}{v(s)}.
    $}
\end{equation}
We highlight that, all the variables are expressed in a coordinate frame attached to the vehicle, where the $x$ axis is the longitudinal direction, the $z$ axis points upwards, normal to the track, and the $y$ axis completes the right-handed coordinate system. We refer to \cite{radrizzani2023optimal} for a pictorial representation. Given the assumptions of point-mass model and fixed trajectory, the lateral dynamics is accounted for via the friction-ellipse static constraint \cite{riva2022sizing,broere20224wheel}, limited by the peak values of longitudinal and lateral friction, $\mu_x$ and $\mu_y$ as:
\begin{equation}
  \left(\frac{F_x}{\mu_x}\right)^2 +\left(\frac{F_y}{\mu_y}\right)^2 \leq F_z^2,
  \label{eq_friction}
\end{equation} 
where total longitudinal, lateral and vertical forces - respectively $F_x$, $F_y$ and $F_z$ - acting on the vehicle are defined as:
\begin{equation}
    \begin{array}{l}
     F_x = \frac{T_w(s)}{R_\mathrm{w}} \\
     F_y = M\rho_R(s) v(s)^2 \\
     F_z = Mg\cos(\theta(s))+F_\mathrm{down}(v(s)),
    \end{array}
   \label{eq_forces}
\end{equation}
where the aerodynamic downforce is expressed as $F_\mathrm{down}=C_\mathrm{down}v(s)^2$, and $C_\mathrm{down}$ is the aerodynamic downforce coefficient.
Exploiting the force definitions, \eqref{eq_friction} becomes
\begin{equation}
\resizebox{1.0\hsize}{!}{$
\left(\frac{T_w(s)}{\mu_x R_w}\right)^2 +\left(\frac{M\rho_R(s) v(s)^2}{\mu_y}\right)^2 \leq \left(Mg\cos({\theta(s)})+C_\mathrm{down}v(s)^2\right)^2.$}
      \label{eq_friction_2}
\end{equation}
It is worth mentioning that, in many literature works, e.g., \cite{ebbesen2018minlap, borsboom2021convex}, the ellipse constraint is relaxed by a rectangle, decoupling longitudinal and lateral limits, allowing the computation in advance of the maximum vehicle speed value achievable along the race-line, combining expressions of lateral and vertical forces in \eqref{eq_forces}. For the purpose of our work, we stick with the ellipse formulation, which represents a better description of the \textit{driving at the limit} concept typical of the racing scenario.\\
From the above description of the vehicle model, we can specify the cost function in \eqref{eq_two_stage_speed}, 
%, as in Tab. \ref{tab_formulation_recap},
the vector of optimization variables $u_v = \left[v(s),T_w(s)\right]$, and the constraint set $\mathscr{V}$,  composed by equations (\ref{eq_initial_condition},\ref{eq_veh_model_space_ref},\ref{eq_friction_2}). In this formulation, which is summarized in Tab. \ref{tab_formulation_recap}, non-convexity arises due to multiple reasons: i) the term $\frac{1}{v}$ in the cost function \eqref{eq_two_stage_speed}, ii) the term $\frac{1}{v}$ and the term proportional to $v^2$ in the state dynamics equality constraint \eqref{eq_veh_model_space_ref}, and iii) the fourth-order polynomial dependence from $v$ in the inequality constraint \eqref{eq_friction_2}.

Nevertheless, different works, starting from \cite{murgovski2015convex} and \cite{ebbesen2018minlap}, have shown that it is possible to rewrite the vehicle dynamics part of the optimization problem into a convex framework. This formulation is achieved from the non-convex one by following the three steps below. 
\begin{itemize}
    \item[(i)] The lethargy $\frac{\mathrm{d}t}{\mathrm{d}s}(s)$ is introduced as an additional optimization variable, making the cost function \eqref{eq_two_stage_speed} linear, thus convex:
    \begin{equation}
        \int^{S}_{0} \frac{1}{v(s)} \mathrm{d}s \rightarrow \int^{S}_{0} \frac{\mathrm{d}t}{\mathrm{d}s}(s) \mathrm{d}s.
    \end{equation}
 Despite this, the link between vehicle speed $v(s)$ and lethargy still introduces the non-convex constraint $v\cdot\frac{\mathrm{d}t}{\mathrm{d}s} = 1$, which can be however relaxed as:
    \begin{equation}
        v(s)\cdot\frac{\mathrm{d}t}{\mathrm{d}s}(s) \geq 1,
        \label{eq_cvx_relax_v}
    \end{equation}
    that is a convex constraint since it can be expressed as the second-order cone (see \cite{borsboom2021convex} and \cite{ebbesen2018minlap} for more details)
    \begin{equation}
        \label{eq_cone_speed_cvx}
                \left|\left| \begin{array}{c}
             2  \\
             v(s)\cdot\frac{1}{\bar{v}}-\bar{v}\cdot\frac{\mathrm{d}t}{\mathrm{d}s}(s)
        \end{array} \right|\right|_2 \leq v(s)\cdot\frac{1}{\bar{v}}+\bar{v}\cdot\frac{\mathrm{d}t}{\mathrm{d}s}(s),
    \end{equation}
    where $\bar{v}=1$ m/s is the normalization speed.
    \item[(ii)] The kinetic energy of the vehicle $E_\mathrm{kin}(s)$ is introduced as a state variable in place of the speed $v(s)$, generating a linear, thus convex, state equation, since the spatial derivative of an energy is a force, and the drag resistance is a linear function of the kinetic energy:
    \begin{equation}
    \begin{aligned}
        \frac{\mathrm{d}E_\mathrm{kin}(s)}{\mathrm{d}s} = &\frac{T_w(s)}{R_w}-\frac{2}{M}\left(C_\mathrm{drag}+C_\mathrm{roll}C_\mathrm{down}\right)E_\mathrm{kin}(s)\\&-Mg\left(\sin(\theta(s))+C_\mathrm{roll}\cos(\theta(s))\right),
        \end{aligned}
        \label{eq_Ekin_cvx}
    \end{equation}
    initialized as:
    \begin{equation}
        E_\mathrm{kin}(0)=\frac{1}{2}M v^2_0.
        \label{eq_Ekin_0}
    \end{equation}
    Finally, the non-convex quadratic equality constraint between kinetic energy and speed, $E_\mathrm{kin}(s) = \frac{1}{2}Mv(s)^2$, is relaxed as: 
    \begin{equation}
    E_\mathrm{kin}(s) \geq \frac{1}{2}Mv(s)^2,
        \label{eq_cvx_relax_Ekin}
    \end{equation}
    to obtain a convex quadratic inequality constraint.
    \item[(iii)] The non-convex ellipse limit in \eqref{eq_friction_2} can be expressed as a convex constraint exploiting the norm operator and the linearity of both centripetal and aerodynamic forces with respect to kinetic energy:
\begin{equation}
    \label{eq_cvx_ellipse}
    \resizebox{0.9\hsize}{!}{$
    \left|\left| \begin{array}{c}
             \frac{T_w(s)}{\mu_x R_w}  \\
             \frac{2\rho_R(s) E_\mathrm{kin}(s)}{\mu_y}
        \end{array} \right|\right|_2 \leq \left(Mg\cos({\theta(s)})+\frac{2}{M}C_\mathrm{down}E_\mathrm{kin}(s)\right)^2
    $}
\end{equation}
\end{itemize}
To retrieve a convex formulation for the vehicle dynamics, the two relaxations \eqref{eq_cvx_relax_v} and \eqref{eq_cvx_relax_Ekin} are necessary, but, fortunately, they do not change the global optimum. Indeed, the solution of the relaxed problem is equivalent to the non-relaxed one, because both constraints would be satisfied with equality. Indeed, 1) to minimize the race time, the lethargy $\leth$ must be minimized, meaning that, given a vehicle speed the first relaxation \eqref{eq_cvx_relax_v} would be satisfied with equality; 2) then, for the same reason, the optimal solution would maximize the vehicle speed, which is limited by the kinetic energy constraint, leading to the tight satisfaction of the second relaxation \eqref{eq_cvx_relax_Ekin}.

To recap, 
the convex formulation of the vehicle dynamics has 
$u_v = \left[\frac{\mathrm{d}t}{\mathrm{d}s}(s),v(s),E_\mathrm{kin}(s), T_w(s)\right]$ as set of optimization variables, and the set $\mathscr{V}$ defined as in Tab. \ref{tab_formulation_recap} as the set of constraints (\ref{eq_cone_speed_cvx},\ref{eq_Ekin_cvx},\ref{eq_Ekin_0},\ref{eq_cvx_relax_Ekin},\ref{eq_cvx_ellipse}).
We claim that the vehicle dynamics part of the optimization problem in \eqref{eq_two_stage_speed} can be expressed, without approximation, as a convex problem, at the price of two additional optimization variables, one for each constraint relaxation, which may increase the computational burden.
We remark also that this convex formulation is not limited to the point-mass model of a vehicle, but it can be extended to more complex scenarios, as shown by \cite{broere20224wheel} with a double-track model for a minimum lap time problem.\\

%% BATTERY
\subsection{Battery Model}
\label{sub_sec_batt}
In this section, we discuss the core topic, namely the modeling of the battery subsystem in Problem \eqref{eq_two_stage_speed}. In a sizing problem, the battery model should be able to capture (i) the internal energy losses, (ii) the dynamics of the disposable energy, and (iii) its physical limitations constraining the achievable performance, while keeping the complexity as low as possible. In this scenario, electro-chemical models \cite{rosewater2019battmodel} are far away from being exploitable, and it is well-established that Equivalent Circuit Models (ECM) can be the right tool to use, see \cite{radrizzani2023optimal,riva2022sizing,vankampen2022maximumdistance,duhr2022convex}. The family of ECMs is wide, and we can search for the right balance between complexity and accuracy, which affects both the structure of the problem and the significance of the results. In the next part of the section, we describe three possible modeling approaches, with increasing complexity, from a non-convex perspective, and then, as in Section \ref{sub_sec_veh}, we discuss how, whether it is possible, to obtain a convex formulation for the subsystem.

All three models share some common variables and equations, that are summarized in this paragraph. The battery pack is made of a multitude of cells, taken from a specific technology. Each cell technology is characterized by its weight $m_\mathrm{cell}$, a nominal capacity $q_\mathrm{cell}$, an internal resistance $r_{\mathrm{0},\mathrm{cell}}$, and an Open-Circuit Voltage (OCV) $v_\mathrm{oc}$, the latter possibly function of the state-of-charge (SoC), labelled with $\zeta$ in the equations, of the cell. From these definitions, the related variables at the battery level can be computed as functions of the battery size parameters, i.e., the number of cells in series $N_s$ and in parallel $N_p$:
\begin{equation}
 V_\mathrm{oc} = N_s v_\mathrm{oc} , \; 
     R_0=\frac{N_s}{N_p} r_{\mathrm{0},\mathrm{cell}} , \;
     Q_b= N_p q_\mathrm{cell}.
   \label{eq_para}
\end{equation}
Each battery model includes also the SoC dynamics, which is expressed through the battery current $I_b(s)$ using the spatial reformulation in \eqref{eq_spt_ref} as:
\begin{equation}
    \frac{\mathrm{d}\mathrm{\zeta}(s)}{\mathrm{d}s} = - \frac{I_b(s)}{Q_b}\cdot \frac{1}{v(s)}.
    \label{eq_soc}
\end{equation}
 Moreover, the whole battery pack weight $M_b$, affecting the vehicle mass $M$ in \eqref{eq_long}, can be expressed as:
\begin{equation}
    \begin{array}{l}
     M_b=  \frac{N_p N_s m_\mathrm{cell}}{\alpha},
    \end{array}
   \label{eq_parM}
\end{equation}
where $\alpha$ is the packaging factor accounting for the weight of the battery case, which is modeled as the $(1-\alpha)$ percentage of the overall weight.
Then, to exploit the battery in its correct operating region, strict constraints, obtained from cell limits on voltage $\left(v_\mathrm{min},v_\mathrm{max}\right)$,  current $\left(i_\mathrm{min},i_\mathrm{max}\right)$, and SoC, are introduced:
\begin{equation}
  \begin{array}{l}
     N_p i_\mathrm{min} \leq I_b(s) \leq N_p i_\mathrm{max} \\
     N_s v_\mathrm{min} \leq V_b(s) \leq N_s v_\mathrm{max} \\
     0 \leq \zeta(s)\leq 1
    \end{array},
  \label{eq_batt_const}
\end{equation}
where $V_{b}$ is the battery voltage at the terminals, whose expression depends on the chosen model. 
Finally, to comply with possible regulations, a limit on the power achievable at the terminals of the battery, i.e., $P_{b}(s)=V_{b}(s)\cdot I_{b}(s)$, is introduced:
\begin{equation}
      P_{b,\min} \leq  V_{b}(s)\cdot I_{b}(s) \leq P_{b,\max}.
  \label{eq_max_batt_pow_const}
\end{equation}
In the following, the peculiarities of each non-convex battery model are described, deriving its own formulation inside the optimization problem. 

%%%%%%%%%%%%%%%%%%%%%%%%%%%
\begin{enumerate}
    \item \textit{Static ECM with constant OCV ($\mathrm{V}_\mathrm{n}$--$\mathrm{R}$).}\\ It is the simplest model, where the open-circuit voltage is assumed to be constant and equal to the nominal value: $V_\mathrm{oc} = N_s v_\mathrm{oc} = N_s v_n= V_n$, where $v_{n}$ is the nominal cell voltage. 
The battery optimization vector is defined as 
$u_b=\left[ \zeta(s), I_{b}(s)\right]$, while the constraint set $\mathscr{B}$, in Tab. \ref{tab_formulation_recap}, contains equations (\ref{eq_soc},\ref{eq_batt_const},\ref{eq_max_batt_pow_const}), where $V_b(s)=V_n - R_0 I_b(s)$.

%%%%%%%%%%%%%%%%%%%%%%%%
\item \textit{Static ECM with SoC dependent OCV ($\mathrm{V}_\mathrm{SoC}$--$\mathrm{R}$)}.\\ This second model integrates the well-known dependence of the open-circuit voltage from the state-of-charge to improve accuracy, i.e., $V_\mathrm{oc}(\zeta) = N_sv_\mathrm{oc}(\zeta)$, where the dependence can be any nonlinear function matching experimental data. In this second model, vector $u_b=\left[ \zeta(s), I_{b}(s)\right]$ is the same as the first one, while the constraint set $\mathscr{B}$ is very similar to the previous case, as shown in Tab. \ref{tab_formulation_recap}, where $V_b(s)= V_\mathrm{oc}(\zeta(s)) - R_0 I_b(s)$.
%%%%%%%%%%%%%%%%%%%%%%%%
\item \textit{Dynamic ECM with SoC dependent OCV ($\mathrm{V}_\mathrm{SoC}$--$\mathrm{RC}$)}.\\
In this last case, we leverage the Randle model, which adds dynamic components to the second  static equivalent circuit in the form of RC pairs, which account for the relaxation dynamics in the cell. In our work, we consider a model with a single RC pair, chosen as the one with the longest time constant \cite{rosewater2019battmodel}, because fast ones are typically characterized by much lower resistance values, thus less impacting on the disposable energy. 
We enlarge the set of optimization variables, $u_b=\left[\zeta(s),I_b(s),V_1(s)\right]$, to include an additional state, namely the voltage $V_1$ across the equivalent RC pair at the battery level, whose resistance and capacitance values at the battery level are obtained using Thevenin's equivalent from cell's ones as $C_1 = c_1\frac{N_p}{N_s}$ and $R_1 = r_1\frac{N_s}{N_p}$, and whose state equation reads as
\begin{equation}
    \label{eq_cap_state}
     \frac{\mathrm{d}V_1(s)}{\mathrm{d}s} = \frac{1}{v(s)}\cdot \frac{1}{C_1}\left( I_b(s) - \frac{V_1(s)}{R_1}\right).
\end{equation}
From this addition, set $\mathscr{B}$ enlarges the previous one including \eqref{eq_cap_state}, as visible in Tab. \ref{tab_formulation_recap}, where $V_b(s)= V_\mathrm{oc}(\zeta(s)) - R_0 I_b(s)-V_1(s)$.\\
\end{enumerate}
%%%%%%%%%%%%%%%%%%%%%%%%
As stated before, all the three introduced models are defined through non-convex constraints, due to the presence of term $\frac{1}{v}$ in the state equations, and, for the last two, of the OCV relationship with respect to the SoC. A convex re-formulation, as demonstrated in \cite{borsboom2021convex,ebbesen2018minlap}, is possible only when the open-circuit voltage is assumed constant and is based on the following steps.
\begin{itemize}
    \item [(i)] Thanks to the constant OCV, battery SoC can be equivalently replaced by the available battery energy $E_b$ as a state variable, which are linked by $E_b(s) = \zeta(s)\cdot Q_b \cdot V_n$. As explained for the vehicle dynamics model, the spatial derivative of energy is a force, meaning that, declaring the open-circuit and terminal forces, $F_\mathrm{oc}(s)$ and $F_b(s)$, from the respective powers $P_\mathrm{oc} (s)$ and $P_\mathrm{b} (s)$ as:
    \begin{equation}
        \label{eq_Foc_Fb_definition}
        \begin{array}{l}
            F_\mathrm{oc}(s) = \frac{P_\mathrm{oc}(s)}{v(s)} = \frac{V_n\cdot I_b(s)}{v(s)} \\
            F_b(s) = \frac{P_{b}(s)}{v(s)} = \frac{(V_n-R_0I_b(s))\cdot I_b(s)}{v(s)},
        \end{array}
    \end{equation}
    we can derive the state equation for the battery energy evolution as:
    \begin{equation}
        \label{eq_batt_energy_evolution_cvx}
        \frac{\mathrm{d}E_b(s)}{\mathrm{d}s} = -F_\mathrm{oc}(s).
    \end{equation}
    \item [(ii)] From simple manipulations of the  equations, the power loss in the internal resistance can be written as:
    \begin{equation}
    \label{eq_cvx_power_loss_battery}
        P_{b,\mathrm{loss}}(s) = P_\mathrm{oc}(s) - P_b(s) = \frac{R_0}{V^2_n}\cdot P^2_\mathrm{oc}(s).
    \end{equation}
    Exploiting the definition of $F_\mathrm{oc}(s)$, $F_b(s)$, and lethargy $\leth$, the power loss in the internal resistance can be written in terms of forces as:
    \begin{equation}
    \label{eq_cvx_force_loss_battery}
        \left(F_\mathrm{oc}(s) - F_b(s)\right)\cdot \frac{\mathrm{d}t}{\mathrm{d}s}(s)  = \frac{R_0}{V^2_n}\cdot F^2_\mathrm{oc}(s).
    \end{equation}
    To obtain a convex formulation, \eqref{eq_cvx_force_loss_battery} is relaxed by:
        \begin{equation}
    \label{eq_cvx_force_loss_battery_relaxed}
\left(F_\mathrm{oc}(s) - F_b(s)\right)\cdot \frac{\mathrm{d}t}{\mathrm{d}s}(s)  \geq \frac{R_0}{V^2_n}\cdot F^2_\mathrm{oc}(s),
    \end{equation}
    which is a second-order cone, as shown in \cite{borsboom2021convex}:
    \begin{equation}
    \label{eq_batt_cone}
    \resizebox{0.9\hsize}{!}{$
    \left|\left| 
    \begin{array}{c}
             2\frac{\sqrt{R_0}}{V_n}F_\mathrm{oc}(s) \sqrt{\frac{\bar{v}}{\bar{F}}}\\
             \left(F_\mathrm{oc}(s)-F_b(s)\right)\cdot \frac{1}{\bar{F}} - \bar{v}\cdot\frac{\mathrm{d}t}{\mathrm{d}s}(s)
    \end{array} 
    \right|\right|_2 
    \leq \left(F_\mathrm{oc}(s)-F_b(s)\right)\cdot\frac{1}{\bar{F}}+\bar{v}\cdot\frac{\mathrm{d}t}{\mathrm{d}s}(s),    $}
    \end{equation}
\end{itemize}
where $\bar{v}=1$ m/s and $\bar{F}=1$ N are normalization terms.
This last relaxation requires an important remark: if on the one hand, it allows to obtain a convex formulation of the model, on the other hand, differently from the vehicle dynamics case, it is satisfied with equality only when the battery energy should be optimized to minimize the race time. In all other cases, the internal battery loss can be overestimated, leading to non-physical profiles, as shown in Fig. \ref{fig:eq_res_convex} for the case study.
Finally, from \eqref{eq_Foc_Fb_definition}, the battery linear inequality constraints \eqref{eq_batt_const} and \eqref{eq_max_batt_pow_const} are rewritten in terms of forces as:
\begin{equation}
  \begin{array}{l}
     N_p i_\mathrm{min} \frac{\mathrm{d}t}{\mathrm{d}s}(s)\leq \frac{F_\mathrm{oc}(s)}{V_n} \leq N_p i_\mathrm{max} \frac{\mathrm{d}t}{\mathrm{d}s}(s)\\
        N_s v_\mathrm{min}  \frac{\mathrm{d}t}{\mathrm{d}s}(s)\leq V_n \frac{\mathrm{d}t}{\mathrm{d}s}(s)-R_0\frac{F_\mathrm{oc}(s)}{V_n} \leq N_s v_\mathrm{max}  \frac{\mathrm{d}t}{\mathrm{d}s}(s)\\
         P_{b,\min} \frac{\mathrm{d}t}{\mathrm{d}s}(s)\leq F_b(s)\leq  P_{b,\max} \frac{\mathrm{d}t}{\mathrm{d}s}(s)\\
     0 \leq E_b(s) \leq  E_b^{\max},
    \end{array},
  \label{eq_batt_const_cvx}
\end{equation}
where $E_b^{\max}=Q_b\cdot V_n$.
From the above description, in the convex reformulation the vector $u_b$ is defined as $u_b=\left[E_b(s), \; F_\mathrm{oc}(s), \; F_b(s)\right]$, while the constraint set $\mathscr{B}$, as in Tab. \ref{tab_formulation_recap}, contains equations (\ref{eq_batt_energy_evolution_cvx},\ref{eq_batt_cone},\ref{eq_batt_const_cvx}).

To conclude, we highlight once more that this convex reformulation is possible only for the model with constant OCV introducing the aforementioned relaxation. Even if in \cite{verbruggen2020joint} a linear dependence of such a relationship is integrated into a convex framework for energy optimization, its applicability can not be extended to the minimum lap-race time problem, due to the needed spatial reformulation.

%% POWERTRAIN AND BRAKING
\subsection{Powertrain, Brakes and Coupling Constraints}
\label{sub_sec_ptr_brk_coupl}
The integration of the battery and vehicle models is achieved via both coupling constraints and powertrain-brakes modeling, whose models are highly influenced by the required level of detail. In racing electric cars, we identify three main components connecting the battery to the vehicle: (i) \textit{inverter}, (ii) \textit{electric motor}, and (iii) \textit{gear-box}. As a preliminary observation, the dynamics involved in these subsystems are typically orders of magnitude faster than the vehicle one, so that, as far as we are aware, no state variables are considered in any previous work (inside vector $u_p$). The only exception regards the temperature dynamics of the electric motor, see \cite{herrmann2020minimum} and \cite{locatello2021time}, which however falls outside the domain of interest of this work, where we stick with the assumption in \cite{riva2022sizing} and \cite{radrizzani2023optimal} of the presence of a properly sized cooling system able to keep the temperature under control.

The \textit{inverter} is rarely accounted for in literature: in \cite{vankampen2022maximumdistance} the losses are modeled through a fitted quadratic function, framing it in the convex scenario through a similar relaxation technique used for the battery pack in \eqref{eq_cvx_force_loss_battery_relaxed}; in\cite{herrmann2022optimization}, instead, the losses are introduced on semiconductor level, including both conduction and switching losses in a non-convex optimization setup.
On the contrary, the \textit{electric motor} is widely considered: \cite{vankampen2022maximumdistance,broere20224wheel} and similarly \cite{borsboom2021convex}, account for the motor limits and its losses via a quadratic model, fitted from data, as functions of the operating point;  \cite{locatello2021time} and \cite{duhr2022convex} employ a simpler model of the losses, just proportional to the square of the output power. In all these approaches, the models of the losses are relaxed, similarly to \eqref{eq_cvx_force_loss_battery_relaxed}, to be included in a convex framework. We remark that the same approaches can be employed without relaxations in a non-convex setup, like in \cite{duhr2023minimum} for F1 hybrid cars.
Finally, regarding the \textit{gear-box}, most literature works, see \cite{vankampen2022maximumdistance} and \cite{broere20224wheel}, use a Fixed Gear Transmission (FGT) model with constant gear ratio and efficiency. However, \cite{borsboom2021convex} enlarges the analysis comparing the FGT model with a Continuously Variable Transmission (CVT), where the losses are included via a fitted quadratic model, always relaxed to be in a convex framework.

Despite more complex models can be included, as in\cite{herrmann2022optimization} and\cite{vankampen2022maximumdistance}, given the energetic perspective of battery sizing problems, we employ the simplest possible one, namely a single average efficiency $\eta$ mapping the entire chain from the battery to the wheel, thus, including the electric motor, the inverter, and the gear-box. In this way, the focus is given to the impact of the different battery models discussed in Section \ref{sub_sec_batt}. Despite its simplicity, 
a fixed efficiency with a bi-directional flow of power can not be easily modeled, since the value of the efficiency is a discontinuous function of the sign of battery current:
\begin{equation}
    T_m(s)\cdot \frac{v(s)}{R_\mathrm{w}} = V_b(I_b(s))\cdot I_b(s) \cdot \eta^{\sign(I_b(s))},
    \label{eq_eta}
\end{equation}
where $T_m(s)$ is the torque delivered by the electric motor, which is part of the powertrain variables $u_p$, together with the brake negative torque $T_\mathrm{br}$: $u_p=\left[T_m(s),T_\mathrm{br}(s)\right]$. The discontinuity in \eqref{eq_eta} can be smoothed as proposed in \cite{riva2022sizing} to obtain a non-convex differentiable equality constraint:
\begin{equation}
\begin{array}{l}
    \eta^{\sign (I_b(s)) } \simeq\\ \hspace{0.5cm} \tilde{\eta}(I_b(s)) = \frac{\eta^{-1}+\eta}{2} + \frac{\eta^{-1}-\eta}{2}\cdot \tanh{(-\beta_\eta I_b(s))}.
    \end{array}
    \label{eq:eta_implementation}
\end{equation}
where $\beta_{\eta}$ represents a design smoothing variable. In conclusion, in the non-convex scenario, the sets $\mathscr{P}$ and $\mathscr{C}$, as shown in Tab. \ref{tab_formulation_recap}, are defined respectively by the constraint on the negative sign of the brake torque \eqref{eq_coupling_noncvx}, and by \eqref{eq_eta} and \eqref{eq:eta_implementation} for the efficiency of the transmission, along with the definition of the total torque at the wheel as the sum of motor and brakes \eqref{eq_coupling_noncvx}.
\begin{equation}
T_w(s) = T_m(s) + T_\mathrm{br}(s) \text{ and } T_\mathrm{br}(s) \leq 0
\label{eq_coupling_noncvx}
\end{equation}

As already stated, the smoothing of the efficiency model in \eqref{eq:eta_implementation} still represents a non-convex constraint. Nevertheless, a convex solution can be obtained employing a technique proposed in the literature for the gear-box efficiency \cite{broere20224wheel}. In this solution, see also Tab. \ref{tab_formulation_recap}, the brake torque does not explicitly appear, so that $u_p$ and set $\mathscr{P}$ can be left empty, since the total torque applied to the wheel $T_w(s)$ includes the electric motor one.
The discontinuous efficiency is accounted for in the coupling constraint via the following double inequality:
\begin{equation}
\label{eq_coupling_cvx}
\left\{\begin{array}{c}
    \frac{T_w(s)}{R_w}-\eta F_b(s) \leq 0\\
    \frac{T_w(s)}{R_w}-\frac{1}{\eta} F_b(s) \leq 0
\end{array}\right.
\end{equation}
The constraint relaxation in \eqref{eq_coupling_cvx} requires a two-fold explanation: (i) $\frac{T_w}{R_w}-\eta F_b \leq 0$ models the traction phase, and it is satisfied with equality when the battery energy should be optimized to minimize the race time; (ii) $\frac{T_w}{R_w}-\frac{1}{\eta} F_b \leq 0$ models the braking phase, where the difference between wheel and efficiency-scaled battery power is allocated to the mechanical brakes as:
\begin{equation}
    T_\mathrm{br}^*(s) = T_w(s)-\frac{R_w}{\eta} F_b(s). 
    \label{eq_brake_torque_eq}
\end{equation}
If, on the one hand, such an implicit brake formulation allows to reduce the number of optimization variables with simpler constraints; on the other hand, it does not allow the simulation of situations where mechanical brakes are absent or power limited, like for the rear emergency brakes of Formula E Generation 3, which can be easily handled with the non-convex formulation in \eqref{eq_coupling_noncvx}.

\begin{table*}[h!]
\caption{Summary of the Minimum Race Time Problem in \eqref{eq_two_stage_speed}. All the models described for vehicle dynamics, battery, and powertrain-brakes are reported, splitting non-convex and convex formulations, whether possible.}
    \label{tab_formulation_recap}
    \centering
    \normalsize
    \resizebox{1\textwidth}{!}{
    \begin{tabular}{ccc}
        \hline
        \rowcolor{gray!50} & \textsc{\textbf{NON-Convex Formulation}} & \textsc{\textbf{Convex Formulation}}  \\
        \hline
        \rowcolor{gray!30}\multicolumn{3}{c}{\textsc{Vehicle Model} } \\
        \hline
        $u_v$ & \multicolumn{1}{l}{\hspace{0.1cm}$\left[v(s), \; T_w(s) \right]$} & \multicolumn{1}{l}{\hspace{0.1cm}$\left[ \frac{\mathrm{d}t}{\mathrm{d}s}(s), \; v(s), E_\mathrm{kin}(s), \; T_w(s)\right]$}\\
        \hline
$\mathscr{V}$ & 
\multicolumn{1}{l}{\hspace{0.1cm}$ 
        \begin{array}{l}
          \frac{\mathrm{d}v(s)}{\mathrm{d}s} = \frac{1}{M}\left( \frac{T_w(s)}{R_w} - \left(C_\mathrm{drag}+C_\mathrm{roll}C_\mathrm{down}\right)v(s)^2-Mg\left(\sin(\theta(s))+C_\mathrm{roll}\cos(\theta(s))\right) \right)\cdot \frac{1}{v(s)}\\
         \left(\frac{T_w(s)}{\mu_x R_w}\right)^2 +\left(\frac{M\rho_R(s) v(s)^2}{\mu_y}\right)^2 \leq \left(Mg\cos({\theta(s)})+C_\mathrm{down}v(s)^2\right)^2\\
         v(0)=v_0
    \end{array}$}
 &  
        \multicolumn{1}{l}{\hspace{0.1cm}$\begin{array}{l}
        \frac{\mathrm{d}E_\mathrm{kin}(s)}{\mathrm{d}s} = \frac{T_w(s)}{R_w}-\frac{2}{M}\left(C_\mathrm{drag}+C_\mathrm{roll}C_\mathrm{down}\right)E_\mathrm{kin}(s)-Mg\left(\sin(\theta(s))+C_\mathrm{roll}\cos(\theta(s))\right)\\
        \frac{1}{2}Mv(s)^2 \leq E_\mathrm{kin}(s)\\
        \left|\left| \begin{array}{c}
             2  \\
             v(s)-\frac{\mathrm{d}t}{\mathrm{d}s}(s)
        \end{array} \right|\right|_2 \leq v(s)+\frac{\mathrm{d}t}{\mathrm{d}s}(s)\\
 \left|\left| \begin{array}{c}
             \frac{T_w(s)}{\mu_x R_w}  \\
             \frac{2\rho_R(s) E_\mathrm{kin}(s)}{\mu_y}
        \end{array} \right|\right|_2 \leq \left(Mg\cos({\theta(s)})+\frac{2}{M}C_\mathrm{down}E_\mathrm{kin}(s)\right)^2\\
        E_\mathrm{kin}(0)=\frac{1}{2}M v^2_0
    \end{array} $ }\\
 \hline
       \rowcolor{gray!30} \multicolumn{3}{c}{\textsc{Battery Model} } \\
        \hline
        \rowcolor{gray!10}\multicolumn{3}{c}{$\mathrm{V}_\mathrm{n}$--$\mathrm{R}$} \\
        \hline
        $u_b$ & \multicolumn{1}{l}{\hspace{0.1cm}$\left[\zeta(s),\; I_b(s)\right]$} & \multicolumn{1}{l}{\hspace{0.1cm}$\left[ E_b(s), \; F_\mathrm{oc}(s), F_b(s)\right]$}\\
        \hline
    $ \mathscr{B}$ & \multicolumn{1}{l}{\hspace{0.1cm}$     \displaystyle    \begin{array}{l}
         \frac{\mathrm{d}\zeta(s)}{\mathrm{d}s} = -\frac{1}{v(s)}\cdot \frac{I_b(s)}{Q_b} \\
         0 \leq \zeta(s) \leq 1\\
         N_p i_\mathrm{min} \leq I_b(s) \leq N_p i_\mathrm{max}\\
         N_s v_\mathrm{min}\leq V_n - R_0I_b(s) \leq N_s v_\mathrm{max}\\
         P_{b,\min}\leq \left( V_n - R_0I_b(s)\right)\cdot I_b(s) \leq P_{b,\max}
    \end{array} $} & \multicolumn{1}{l}{\hspace{0.1cm}$ \displaystyle
        \begin{array}{l}
        \frac{\mathrm{d}E_b(s)}{\mathrm{d}s} = -F_\mathrm{oc}(s) \\
        \left|\left| \begin{array}{c}
             2 \frac{\sqrt{R_0}}{V_n} F_\mathrm{oc}(s)\\
             F_\mathrm{oc}(s)-F_b(s)-\frac{\mathrm{d}t}{\mathrm{d}s}(s)
        \end{array} \right|\right|_2 \leq F_\mathrm{oc}(s)-F_b(s)+\frac{\mathrm{d}t}{\mathrm{d}s}(s)\\
        N_p i_\mathrm{min} \frac{\mathrm{d}t}{\mathrm{d}s}(s)\leq \frac{F_\mathrm{oc}(s)}{V_n} \leq N_p i_\mathrm{max} \frac{\mathrm{d}t}{\mathrm{d}s}(s)\\
        N_s v_\mathrm{min} \frac{\mathrm{d}t}{\mathrm{d}s}(s)\leq V_n \frac{\mathrm{d}t}{\mathrm{d}s}(s)-R_0\frac{F_\mathrm{oc}(s)}{V_n} \leq N_s v_\mathrm{max} \frac{\mathrm{d}t}{\mathrm{d}s}(s)\\
         P_{b,\min} \frac{\mathrm{d}t}{\mathrm{d}s}(s)\leq F_b(s)\leq  P_{b,\max} \frac{\mathrm{d}t}{\mathrm{d}s}(s)\\
        0 \leq E_b(s) \leq  E_b^{\max}\\
    \end{array}$}\\
     \hline
    \rowcolor{gray!10}\multicolumn{3}{c}{$\mathrm{V}_\mathrm{SoC}$--$\mathrm{R}$}\\
    \hline
        $u_b$ & \multicolumn{1}{l}{\hspace{0.1cm}$\left[\zeta(s),\; I_b(s)\right]$} & \\
    \hline
        $\mathscr{B}$ & \multicolumn{1}{l}{\hspace{0.1cm}$ 
        \begin{array}{l}
         \frac{\mathrm{d}\zeta(s)}{\mathrm{d}s} = -\frac{1}{v(s)}\cdot \frac{I_b(s)}{Q_b} \\
         0 \leq \zeta(s) \leq 1\\
         N_p i_\mathrm{min} \leq I_b(s) \leq N_p i_\mathrm{max}\\
         N_s v_\mathrm{min}\leq V_\mathrm{oc}(\zeta(s)) - R_\mathrm{0
         }I_b(s) \leq N_s v_\mathrm{max}\\
         P_{b,\min}\leq \left( V_\mathrm{oc}(\zeta(s)) - R_0I_b(s)\right)\cdot I_b(s) \leq P_{b,\max}
    \end{array} $ }& \\
      \hline
    \rowcolor{gray!10}\multicolumn{3}{c}{$\mathrm{V}_\mathrm{SoC}$--$\mathrm{RC}$}\\
    \hline
        $u_b$ & \multicolumn{1}{l}{\hspace{0.1cm}$\left[\zeta(s),\; V_1(s), \;I_b(s)\right]$} & \\
    \hline
    $\mathscr{B}$ & \multicolumn{1}{l}{\hspace{0.1cm}$
    \begin{array}{l}
         \frac{\mathrm{d}\zeta(s)}{\mathrm{d}s} = -\frac{1}{v(s)}\cdot \frac{I_b(s)}{Q_b} \\
     \frac{\mathrm{d}V_1(s)}{\mathrm{d}s} = \frac{1}{v(s)}\cdot \frac{1}{C_1}\left( I_b(s) - \frac{V_1(s)}{R_1}\right) \\
         0 \leq \zeta(s) \leq 1\\
         N_p i_\mathrm{min} \leq I_b(s) \leq N_p i_\mathrm{max}\\
         N_s v_\mathrm{min}\leq V_\mathrm{oc}(\zeta(s)) - R_\mathrm{0
         }I_b(s) -V_1(s) \leq N_s v_\mathrm{max}\\
         P_{b,\min}\leq \left( V_\mathrm{oc}(\zeta(s)) - R_0I_b(s)-V_1(s)\right)\cdot I_b(s) \leq P_{b,\max}
    \end{array}$} & \\
    \hline
    \rowcolor{gray!30}\multicolumn{3}{c}{\textsc{Powertrain, Brakes and Coupling Constraints}}\\
    \hline
        $u_p$ & \multicolumn{1}{l}{\hspace{0.1cm}$\left[T_m(s),\; T_\mathrm{br}(s) \right]$} & \multicolumn{1}{l}{\hspace{0.1cm}} \\
    \hline
    $\mathscr{P}$ & \multicolumn{1}{l}{\hspace{0.1cm}$T_\mathrm{br} \leq 0$} &  \multicolumn{1}{l}{\hspace{0.1cm}}\\
    \hline
    $\mathscr{C}$ & \multicolumn{1}{l}{\hspace{0.1cm}$\begin{array}{l}
         T_w(s) = T_m(s) + T_\mathrm{br}(s) \\
        T_m(s)\frac{v(s)}{R_\mathrm{w}} = V_b(I_b(s))\cdot I_b(s) \cdot \tilde{\eta}(I_b(s))
        \end{array}$} &  \multicolumn{1}{l}{\hspace{0.1cm}$\begin{array}{l}
    \frac{T_w(s)}{R_w}-\eta F_b(s) \leq 0\\
    \frac{T_w(s)}{R_w}-\frac{1}{\eta} F_b(s) \leq 0\\
\end{array}$}\\
\hline
\rowcolor{gray!30}\multicolumn{3}{c}{\textsc{Cost Function}}\\
    \hline
    $t_\mathrm{race}$ & \multicolumn{1}{l}{\hspace{0.1cm}$\displaystyle \int^S_0 \frac{1}{v(s)} \mathrm{d}s$} & \multicolumn{1}{l}{\hspace{0.1cm}$\displaystyle \int^S_0 \frac{\mathrm{d}t}{\mathrm{d}s}(s) \mathrm{d}s$}\\
    \hline
    \end{tabular}
    }
\end{table*}

\section{The Impact of Battery Model \\ on MRT and Sizing: A Case Study}
\label{sec:comparison}
To evaluate the impact of the different modeling choices described in Section \ref{sec:modelling}, a case study is addressed. Similarly to \cite{riva2022sizing,radrizzani2023optimal}, the vehicle is a Formula E car, specifically the Generation 3, and its parameters are shown in Tab. \ref{tab_vehicle_param}.
\begin{table}[t]
\centering
\caption{Vehicle Parameters}
\begin{tabular}{cccc}
\hline
\textbf{vehicle mass} & $M_v$ & 426 & kg \\
\textbf{wheel radius} & $R_w$ & 34.54 & cm \\
\textbf{drag coefficient} & $C_\mathrm{drag}$ & 0.3927 & Ns\textsuperscript{2}/m\textsuperscript{2} \\
\textbf{aerodynamic downforce coefficient} & $C_\mathrm{down}$ & 0.9526 & Ns\textsuperscript{2}/m\textsuperscript{2} \\
\textbf{rolling coefficient} & $C_\mathrm{roll}$ & 0.015 & - \\
\textbf{friction coefficients} & $\mu_x,\mu_y$ & 1.2 & -\\
\textbf{maximum voltage} & $V_\mathrm{max}$ & 878 & V\\
\textbf{powertrain efficiency} & $\eta$ & 0.87 & -\\
\textbf{packaging factor} & $\alpha$ & 0.80 & -\\
\textbf{maximum traction battery power} & $P_{b,\max}$ & 350 & kW\\
\textbf{maximum braking battery power} & $P_{b,\min}$ & -600 & kW\\
\hline
\end{tabular}
\label{tab_vehicle_param}
\end{table}
Concerning the battery cell, we considered a single technology, the Sony Murata US18650VTC6 (VTC6), representative of the racing context, as demonstrated by its usage in the Formula E Generation 2 championship. Its parameters, are reported in Tab. \ref{tab_vtc6_param}, highlighting if they are available from easy-access datasheets\footnote{www.murata.com/en-eu/products/batteries/cylindrical\label{fn1}}\textsuperscript{,}\footnote{www.imrbatteries.com/content/sony\_us18650vtc6-2.pdf\label{fn2}} or if an identification process is required. The identified parameters come from three different identification procedures proposed in \cite{estaller2022vtc6}, considered as possible alternative models. To summarize, the following models are considered:
\begin{itemize}
    \item[(i)] \textbf{$\mathrm{V}_\mathrm{n}$--$\mathrm{R}$}, characterized by the nominal constant open-circuit voltage and one resistance;
    \item[(ii)] \textbf{$\mathrm{V}_\mathrm{SoC}$--$\mathrm{R}$}, with a SoC-dependent open-circuit voltage and one resistance;
    \item[(iii)] \textbf{$\mathrm{V}_\mathrm{SoC}$--$\mathrm{RC}$}, with an RC couple in addiction to $\mathrm{V}_\mathrm{SoC}$-$\mathrm{R}$, taking into account the three sets of values in Tab. \ref{tab_vtc6_param}.
\end{itemize}

\begin{table}[t]
\centering
\caption{Battery Cell Parameters}
\begin{tabular}{cccc}
\hline
\multicolumn{4}{l}{\textit{from datasheets}\footref{fn1}\textsuperscript{,}\footref{fn2}} \\
\textbf{cell capacity} & $q_\mathrm{cell}$ & 3 & Ah  \\
\textbf{cell mass} & $m_\mathrm{cell}$ & 46.6 & g  \\
\textbf{nominal voltage} & $v_{n}$ & 3.6 & V \\
\textbf{voltage range} & $v_\mathrm{min},v_\mathrm{max}$ & 2, 4.2 & V \\
\textbf{current range} & $i_\mathrm{min},i_\mathrm{max}$ & -6, 30 & A \\
\textbf{resistance} & $r_0$ & 13 & m$\Omega$  \\
\textbf{OCV} & \multicolumn{3}{c}{see \footref{fn1}} \\
\multicolumn{3}{c}{}  \\
& \multicolumn{3}{c}{\includegraphics[scale = 0.9]{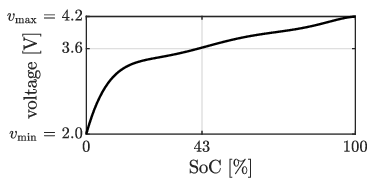}}  \\
% \hline
\multicolumn{4}{l}{\textit{identified in \cite{estaller2022vtc6}}}\\
%\hline
\textbf{RC set 1} & $r_1,c_1$ & 2.28, 4047.82 & m$\Omega$,F   \\
\textbf{RC set 2} & $r_1,c_1$ & 15.04, 995.59 & m$\Omega$,F   \\
\textbf{RC set 3} & $r_1,c_1$ & 20.65, 1344.85 & m$\Omega$,F   \\
\hline
\end{tabular}
\label{tab_vtc6_param}
\end{table}

The impact of the modeling strategies is analyzed following these two steps: (i) the optimal battery size -- for the 23 laps of the 2021 Rome ePrix (see \cite{radrizzani2023optimal} for a pictorial representation) -- is compared using the different battery models in the non-convex formulation; (ii) the evolution of vehicle speed and battery variables is discussed for the optimal battery size, highlighting the potential differences induced by the different battery models; finally, the consequences of using the convex formulation are also discussed. We highlight that, in this specific scenario, the slope $\theta(s)$ is set equal to zero, because the considered track is flat. 
Both approaches have been discretized with $\mathrm{d}s=15$ m using the fourth-order explicit Runge-Kutta method \cite{butcher1987rungekutta}, which keeps the linearity -- and so convexity -- in the convex formulation, and increases the overall accuracy in presence of significant discretization lengths. The value of the discretization step, coherent with previous works in \cite{riva2022sizing} and \cite{radrizzani2023optimal}, is chosen as a compromise between accuracy and complexity, allowing to achieve both consistent sizing results and speed profiles, even when compared with finer space discretizations. The non-convex approach is parsed with CasADi \cite{andersson2019casdi} and solved by employing the interior-point algorithm (IPOPT)  \cite{byrd1999ipopt}  combined with the MA57 linear solver \cite{duff2004ma57}. The convex approach, instead, is parsed with YALMIP \cite{lofberg2004YALMIP} and solved using MOSEK \cite{andersen2000mosek}. Regarding the computational load, we highlight that there are no strict requirements since the optimizations are performed offline. To a give a flavour, the non-convex case with the  $\mathrm{V}_\mathrm{n}$--$\mathrm{R}$ battery model, which has a total of 30362 optimization variables and 75903 constraints, takes on average 43.5 s to return the optimal solution on a personal computer equipped with a 11th Gen Intel(R) Core(TM) i7-11800H @ 2.30GHz   2.30 GHz processor and a 16GB RAM.

\subsection{The Impact of Battery Models}
\label{subsec_impact_battery_model}
\begin{figure}[h]
    \centering
    \includegraphics[width = 0.9\columnwidth]{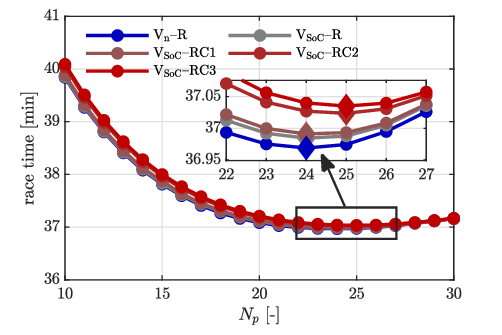}
    \caption{Comparison of sizing results for all the discussed non-convex battery models, as a function of the number of parallels $N_p$. A zoom around the optimum highlights the similarity among the obtained results.}
    \label{fig:sizing_model_sens}
\end{figure}

The cost function to be minimized, i.e., the race time, in this case study is a sole function of the number of cells in parallel $N_p$, which is chosen as the battery size design parameter: $u_s = N_p$. Indeed, the battery cell technology is fixed to be the VTC6, while the number of cells in series $N_s$ is 209, so to impose the maximum battery voltage: $N_s v_\mathrm{max} = V_\mathrm{max}$. Therefore, to compute the solution, the race time is evaluated for $N_p \in \{10,...,30\}$ with a grid approach. Results are shown in Fig. \ref{fig:sizing_model_sens}: it is visible that the computed race time increases along with the model complexity. This trend is motivated by the reduction of net battery capacity, due to both the nonlinear open-circuit voltage characteristics and the total resistance increase ($R_0+R_1$). We highlight that the set of RC couples is intentionally sorted with increasing $R_1$ in Tab. \ref{tab_vtc6_param} to enhance the readability of the figure. Concerning the optimal point, the first three models ($\mathrm{V}_\mathrm{n}$--$\mathrm{R}$, $\mathrm{V}_\mathrm{SoC}$--$\mathrm{R}$, $\mathrm{V}_\mathrm{SoC}$--$\mathrm{RC1}$) share the same optimum $N_p=24$, while the other two ($\mathrm{V}_\mathrm{SoC}$--$\mathrm{RC2}$, $\mathrm{V}_\mathrm{SoC}$--$\mathrm{RC3}$) have the optimal point in $N_p=25$. Indeed, this is a natural consequence of a lower net capacity, which calls for a bigger battery size. However, looking at the negligible race time difference between the solutions with $24$ and $25$ cells in parallel for the more complex models, we can appreciate the limited error introduced by a simpler battery model.

\begin{figure}[t]
    \centering
    \includegraphics[width = 0.9\columnwidth]{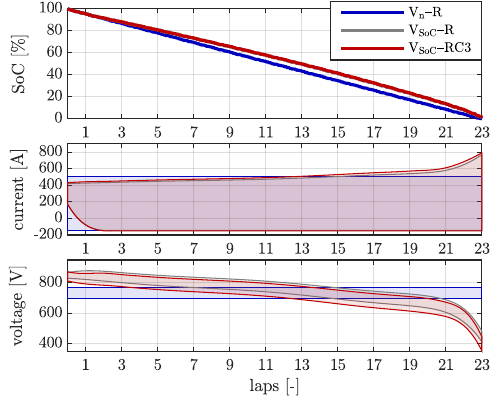}
    \caption{Comparison of SoC, current and voltage race profiles with $N_p =24$, for the three battery models with increasing complexity. For clarity, only the operating region envelopes are depicted for battery currents and voltages.}
    \label{fig:race_profiles}
\end{figure}

In order to compare the evolution of different variables during the race, we consider now $N_p = 24$. Fig. \ref{fig:race_profiles} compares the behavior, in the lap-domain, of both SoC and operating regions of battery current and voltage for $\mathrm{V}_\mathrm{n}$--$\mathrm{R}$, $\mathrm{V}_\mathrm{SoC}$--$\mathrm{R}$ and $\mathrm{V}_\mathrm{SoC}$--$\mathrm{RC3}$; the latter being the model with RC values with the highest impact on the results (see again Fig. \ref{fig:sizing_model_sens}). All three solutions end the race with almost zero state of charge, leaving the battery without available energy, meaning that the battery is not oversized. We can also see that the $\mathrm{V}_\mathrm{n}$--$\mathrm{R}$ state of charge profile is more different with respect to $\mathrm{V}_\mathrm{SoC}$--$\mathrm{R}$ and $\mathrm{V}_\mathrm{SoC}$--$\mathrm{RC3}$, which are closer to each other. This aspect is motivated by the real shape of the OCV in Tab. \ref{tab_vtc6_param} that is neglected in $\mathrm{V}_\mathrm{n}$--$\mathrm{R}$, using the constant nominal one. This has a consequent impact on the battery current and voltage operating regions. At the beginning of the race, the battery is fully charged, meaning that the open-circuit voltage is close to the maximum value allowed at the terminals (as visible from the picture in Tab. \ref{tab_vtc6_param}), thus limiting the maximum negative current achievable during braking. On the contrary, when the SoC lowers, the open-circui voltage decreases as well, so that the battery current should increase during traction phases to match the maximum power constraint \eqref{eq_max_batt_pow_const}.

\begin{figure}[h!]
    \centering
    \includegraphics[width = 0.9\columnwidth]{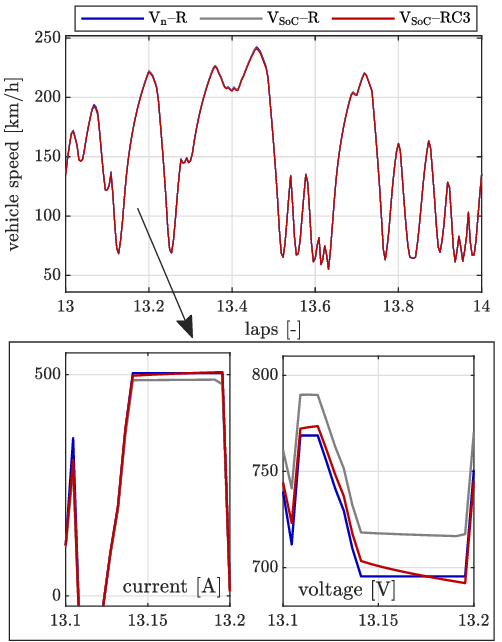}
    \caption{Comparison of vehicle speed, in the fourteenth lap of the race, with $N_p =24$, for the three battery models with increasing complexity, a zoom of the current and voltage is given, to appreciate the difference between the models.}
    \label{fig:lap_profiles}
\end{figure}

Focusing on the fourteenth lap in Fig. \ref{fig:lap_profiles}, it is possible to compare in more detail the different modeling approaches, because, in this lap, the open-circuit voltages of the more complex models are close to the nominal one. First of all, we can appreciate the similarity among the optimized speed profiles, despite the differences in the voltage generated by the RC dynamics in $\mathrm{V}_\mathrm{SoC}$--$\mathrm{RC3}$, which are compensated by tiny current dissimilarities to match the power requirement. These comments naturally reflect on the achieved race times, as evident by the alike shapes in Fig. \ref{fig:sizing_model_sens}.
The reason behind that is something specific for the racing scenario, where the battery power request is characterized by quick changes from hard braking to full traction, which are much faster than the RC dynamics. Indeed, considering the RC set 3, the time constant $R_1 C_1$ is equal to $27.77$ s, while the change between traction and braking occurs approximately every 3-4 seconds.

\subsection{The Impact of Convex Formulations}
\label{subsec_convex_vs_nonconvex}
\begin{figure}[t]
    \centering
    \includegraphics[width = 0.9\columnwidth]{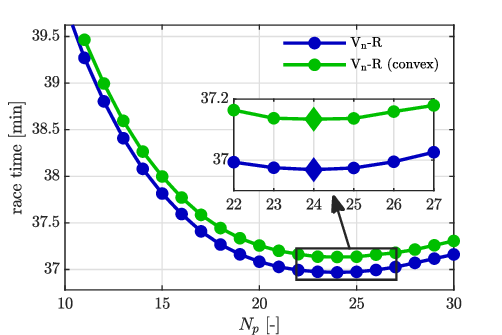}
    \caption{Comparison of sizing results, as a function of the number of parallels $N_p$, between the convex and non-convex approaches for the simplest battery model ($\mathrm{V}_\mathrm{n}$--$\mathrm{R}$).}
    \label{fig:sizing_cvx_vs_nocvx}
\end{figure}

\begin{figure}[t]
    \centering
    \includegraphics[width = 0.9\columnwidth]{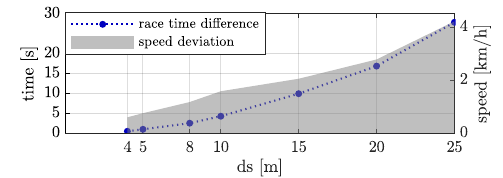}
    \caption{Comparative analysis of the integration errors between the convex and non-convex formulation of the simplest battery model ($\mathrm{V}_\mathrm{n}$--$\mathrm{R}$). Race time discrepancies and distributions of vehicle speed differences are depicted as function $\mathrm{d}s$.}
    \label{fig_cvx_ncvx_ds_vx}
\end{figure}

The analysis now focuses on the convex formulation, comparing it with 
the non-convex one with a static battery model with nominal voltage ($\mathrm{V}_\mathrm{n}$--$\mathrm{R}$), because as discussed in Section \ref{sec:modelling}, they are equal except for the relaxation of the constraints (\ref{eq_cvx_relax_v},\ref{eq_cvx_relax_Ekin},\ref{eq_cvx_force_loss_battery_relaxed},\ref{eq_coupling_cvx}). Indeed, Fig. \ref{fig:sizing_cvx_vs_nocvx} shows that the two approaches have the same optimal solution, and the cost function only differs due to numerical errors in the integration, induced by solving two different formulations of the same problem, where different state equations are employed. 
The motivation of this statement is provided in Fig. \ref{fig_cvx_ncvx_ds_vx}, where the difference between the two approaches is analyzed with respect to the discretization step $\mathrm{d}s$, for $N_p = 24$. In the figure, the gap between achieved race times and the distributions of vehicle speed differences along the race are reported. We can highlight the monotonic trend towards zero of both quantities when the discretization step decreases, providing thorough evidence that both solutions converge towards the same continuous time formulation of the problem.

\begin{figure}[t]
    \centering
    \includegraphics[width = 0.89\columnwidth]{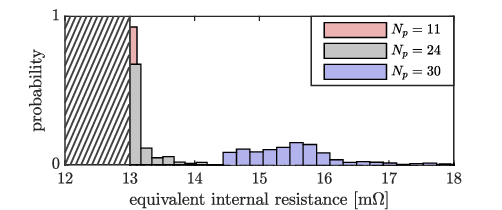}
    \caption{Distributions of equivalent battery internal resistance along the race for three different battery sizes of the convex problem.}
    \label{fig:eq_res_convex}
\end{figure}

\begin{figure}[t]
    \centering
    \includegraphics[width = 0.89\columnwidth]{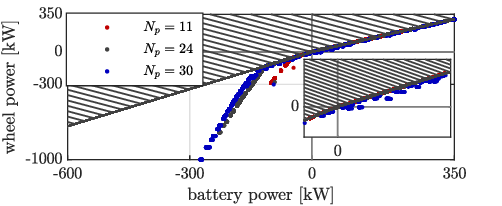}
    \caption{Comparison of the equivalent powertrain efficiency in the battery-wheel power plane for three different battery sizes of the convex problem.}
    \label{fig:eq_eff_convex}
\end{figure}

The aforementioned relaxations can have an important impact on the behavior of the time-lap varying variables. Indeed, if it is proven in \cite{ebbesen2018minlap}, and previously discussed in Section \ref{sub_sec_veh}, that to minimize the race time the relaxed constraints (\ref{eq_cvx_relax_v},\ref{eq_cvx_relax_Ekin}) are satisfied with equality, this is not guaranteed for the other two constraints (\ref{eq_cvx_force_loss_battery_relaxed},\ref{eq_coupling_cvx}).
Starting with the relaxation of the battery constraint, we can compute in post-processing the value of an equivalent internal resistance $R_0^*(s)$ that satisfies the constraint \eqref{eq_cvx_force_loss_battery_relaxed} with equality:
\begin{equation}
    R_0^*(s) = V_n^2 \cdot (F_\mathrm{oc}(s)-F_b(s)) \cdot \leth (s) \cdot \frac{1}{F^2_\mathrm{oc}(s)}.
\end{equation}
The distribution of the values of $R_0^*(s)$ is shown in Fig. \ref{fig:eq_res_convex} for three battery packs with a different number of cells in parallel. It is visible that: if the battery is oversized ($N_p = 30$), the average value of $R_0^*(s)$ is higher than $R_0$, because some energy can be freely dissipated without affecting the race time; on the contrary, when the battery is undersized ($N_p = 11$), $R_0^*(s) = R_0$ because all the energy is necessary to minimize the race time to finish the race. Finally, we can notice that when $N_p = 24$, the equivalent resistance is closer to the real one $R_0$, because the battery is optimally sized to cover the race on the considered circuit.\\
Similar considerations can be made for the relaxation of the powertrain constraint \eqref{eq_coupling_cvx}. Fig. \ref{fig:eq_eff_convex} shows the equivalent efficiency of the powertrain in the battery-wheel power plane, where the fixed efficiency and the forbidden region are highlighted. During traction phases (positive battery power), the optimal solution sticks to the desired fixed efficiency when the battery is either optimally sized or downsized, while it can introduce undesired dissipation when the battery is oversized, for the same reasons as before. In braking phases (negative battery power), the optimal solution is correctly not on the fixed efficiency curve, since the difference is dissipated through the mechanical brakes, as already discussed in Section \ref{sec:modelling}.

These last discussions allow us to better highlight the possible drawbacks of the convex formulations for the minimum race time problem. First of all, a convex formulation is possible only for the simplest battery model, with constant OCV, thus limiting the accuracy and the capability to describe the different stages of the race. Moreover, the need of constraint relaxation to achieve convexity might lead to non-physical behaviors in the optimal profiles of energy-related variables, which turn out to be less informative for further analyses and developments. 
We remark here that, with the specific powertrain layout considered in this work, it is still possible to recompute the electrical variables, like battery power and current, in post-process, moving backwards from the wheel power through the non-relaxed models. As a final comment, such post-processing might not always be applied. For instance, with the hybrid battery pack layout in \cite{radrizzani2023optimal}, an explicit model to split the total battery power between the two sources would not be available.

\subsection{Final Discussion and Sizing Guidelines} 
\label{subsec_casestudy_discussion}
From the case study, we can retrieve important information about the minimum race time optimization and the optimal battery sizing for an electric racing car: if, on the one hand, more complex models are fundamental to compute the real evolution of the control variables, on the other hand, simpler ones are sufficient for sizing purposes. As a matter of fact, we have shown how the problem can be simplified up to a convex formulation without experiencing significant errors in the sizing results, as demonstrated by the negligible differences among the different battery models. As a consequence, we can state that the model with constant OCV attains consistent results and can be used as a reference for this purpose, even though the obtained profiles of electrical variables might be not physical.
 Moreover, since we have shown in Section \ref{subsec_convex_vs_nonconvex} that only small race time differences, due to integration errors, are present between its convex and non-convex formulations, we can stick with the former one, exploiting its global optimality properties.

Summarizing the results of this case study, they suggest a guideline on how to approach the battery sizing problems in racing vehicles: \begin{enumerate}
    \item[(i)] optimal sizing can be formulated with the simplest model; in this way, cell information can be retrieved from easy-access datasheets, which typically report the cell weight, the value of the internal resistance, the nominal voltage and the cell operating limits;
    \item[(ii)] more complex models can be experimentally identified, once a cell technology has been chosen and the battery pack sized;
    \item[(iii)] these models can be employed in a minimum race time problem in order to understand more trustworthy evolutions of the variables during the race, with the purpose of designing real-time control strategies.
\end{enumerate}

\section{Conclusions}
\label{sec:conclusions}
In this work, we reviewed the minimum lap-race time literature approaches for racing electric vehicles, with a focus on their application to sizing. Through the definition of a general mathematical framework, we analyzed the impact of different modeling choices on the resulting optimization problem, with an eye on convexity. In particular, we focused on different battery models, ranging from static electro-equivalent models with nominal voltage, to dynamical equivalent circuits with a SoC-dependent open-circuit voltage. 
Their impact on the battery sizing problem has been evaluated through a case study. In this way, it was possible to show that, when the minimum race time paradigm is employed for sizing purposes, simple convex models are more than enough.
On the contrary, it is also shown how more complex battery models, requiring non-convex formulations, are more suitable to understand and analyze the time evolution of battery and vehicle variables for control purposes.

Future works will focus on extending the analysis towards the other main components of the electric powertrain: electric motors, transmissions and power electronics.

%%%%%%%%%%%%%%%%%%%%%%%%%%%%%%%%%%%%%%%%%%%%%%%%%%%%%%%%%%%%%%%%%%%%%%%%%%%%%%%%
% BIBLIOGRAPHY
%%%%%%%%%%%%%%%%%%%%%%%%%%%%%%%%%%%%%%%%%%%%%%%%%%%%%%%%%%%%%%%%%%%%%%%%%%%%%%%%
\bibliographystyle{IEEEtran}
\bibliography{biblio}

\end{document}